**Title:**

# Harnessing Interpretable Machine Learning for Holistic Inverse Design of Origami

**Authors:**

Yi Zhu[1,2*], Evgueni T. Filipov[1,2]

1. Department of Civil and Environmental Engineering, University of Michigan at Ann Arbor
2. Department of Mechanical Engineering, University of Michigan at Ann Arbor

*. Corresponding Author (email: yizhucee@umich.edu)

## Abstract

This work harnesses interpretable machine learning methods to address the challenging inverse design problem of origami-inspired systems. We show that a decision tree-random forest method is particularly suitable for fitting origami databases, containing both design features and functional performance, to generate human-understandable decision rules for the inverse design of functional origami. First, the tree method is unique because it can handle complex interactions between categorical features and continuous features, allowing it to compare different origami patterns for a design. Second, this interpretable method can tackle multi-objective problems for designing functional origami with multiple and multi-physical performance targets. Finally, the method can extend existing shape-fitting algorithms for origami to consider non-geometrical performance. The proposed framework enables holistic inverse design of origami, considering both shape and function, to build novel reconfigurable structures for various applications such as metamaterials, deployable structures, soft robots, biomedical devices, and many more.

**Introduction:**

Origami, the art of folding paper, provides a method to build novel 3D engineering structures from flat 2D surfaces [1, 2]. These origami structures can be used in a variety of applications such as biomedical devices [3, 4], micro/soft robots [5, 6, 7], frequency selective surfaces [8, 9], metamaterials [10, 11], aerospace structures [12, 13], and many more. Over the years, there have been a number of inverse design methods proposed for origami systems [14, 15, 16, 17], but these methods and algorithms only solve kinematic design problems like fitting origami to arbitrary shapes and geometries. Designing functional origami structures for general engineering applications is still difficult because these active origami systems can have highly nonlinear motions, variable properties, and unintuitive multi-physical behaviors. Properly addressing a generic inverse design problem of origami requires considering the interaction between categorical and continuous features and handling formulations with multiple multi-physical objectives. These problems are difficult to handle with existing optimization-based inverse design methods for origami shape fitting so new solutions are needed.

Machine learning has proven to be a versatile and powerful method to solve physical science problems [18], financial problems [19], biomedical problems [20], E-sport games [21], etc. Moreover, a large number of different machine learning methods like neural network [22], rule list [23], boosting [24], random forest [25], and others have been developed to solve problems of different size, complexity, and nature. Because of the broadness and diversity of these methods, one key challenge on using machine learning is to select the appropriate method for a given problem. For origami type problems, machine learning techniques have been used to predict chaotic dynamic responses [26] and to solve for origami folding motions [27]. However, no prior work has tackled the inverse design problem for origami using machine learning techniques.

In this work, we show that an interpretable machine learning method called the decision tree method and its ensemble version called random forest [25, 28] are particularly suitable for the inverse design of functional origami. Figure 1 summarizes the fundamental idea of this work. The design of origami can be thought of as building a nonlinear function $f$ to calculate the *performance indices* of the system (such as stiffness, Poisson's ratio, material cost, etc.) based on given *design features* (such as the number of origami cells, the thickness of materials, sector angles of the origami pattern, etc.). Usually, numerical simulations are used to represent this function $f$ because it is too complex to be expressed in a closed form. In a traditional setting, designing origami structures is accomplished through plotting and observing the relationships between the features and the performance using nonlinear simulation methods [29, 30]. To extend this traditional design method to an inverse design setup, one needs to construct an *inverse relationship* $f^{-1}$ that calculates a set of design features from the target performance. In this work, we show that it is possible to compute this inverse relationship $f^{-1}$ by first populating an origami performance database and then applying interpretable machine learning to fit the database (approximating $f$). Unlike standard "black box" machine learning methods, interpretable machine learning can produce human-understandable decision rules with which people can understand why certain judgments are made by the algorithms [31, 32]. Once an *interpretable* approximation of the nonlinear function $f$ is obtained, the inverse relationship $f^{-1}$ can be constructed. Through our implementation and exploration of the decision tree method for functional origami systems, we found that selecting the "more informative" branches provides useful design rules for this set of problems (finding the inverse relationship $f^{-1}$).

In the following sections, we demonstrate how to resolve the origami inverse design problem as a data science and machine learning problem. We show how to build the origami performance database and use the decision tree-random forest method to compute human-understandable decision rules for the inverse design of functional origami. First, we demonstrate the methodology using a simple design problem on Miura origami cells. Next, we show that the proposed method can handle categorical features and compare different origami patterns for a more complex origami metasheet design problem. We then show that the tree-based method can tackle multi-objective and multi-physical problems by designing a set of electro-thermal origami grippers. Finally, with a design problem on origami arches, we demonstrate that the proposed method can enable origami shape fitting algorithms to further consider non-geometrical properties so that holistic origami design can be achieved. These examples show that the decision tree-random forest method offers unprecedented versatility for the inverse design of origami systems.

**Results:**

**Computing Origami Design Rules with Interpretable Machine Learning**

This section introduces how the decision tree-random forest method can be used to compute decision rules for the inverse design of origami systems. To demonstrate the methodology, we use a simple design problem on Miura-ori unit cells as presented in Fig. 2A. The design objective is to find a Miura-ori unit cell to have an axial stiffness $k$ smaller than $6000\ N/m$ (target performance: $k < 6000\ N/m$). The Miura-ori cell is cut from a square sheet and is folded to a 60% extension ratio, defined using the ratio between the folded length and flat length of the pattern ($Ext = L'/L = 60\ \%$). The axial stiffness is calculated by fixing the unit cell at one end and applying small forces to stretch it at the other end. For this problem, four design variables (features) are used including the thickness of panels ($t_p$), the thickness of creases ($t_c$), the width of creases ($W$), and the sector angle of the Miura-ori pattern ($\gamma$). We determine the design ranges for these features based on practical fabrication and material limits, and they are: $1.0\ mm < t_p < 6.0\ mm$, $0.5\ mm < t_c < 1.0\ mm$, $1.0\ mm < W < 4.0\ mm$, and $50° < \gamma < 80°$.

To build a database for training the machine learning model, we randomly create designs using different feature values within the specified ranges (step 1 in Fig. 2B). Each column of the table in Fig. 2B represents one design (data point) and 1000 random designs are generated for this example problem. Next, we compute the axial stiffness (or the performance) for each data point using a physics-based origami simulation package called SWOMPS [33] (step 2 in Fig. 2B). The SWOMPS package is used in this paper because it is computationally efficient and can capture multi-physical behaviors important for many origami designs [34, 35] (see Materials and Methods). Based on the simulated performance of each data point, we assign class labels depending on if the design can achieve the target performance or not (step 3 in Fig. 2B). Data points that meet the target ($k < 6000\ N/m$) are assigned a label of 1 while the remaining samples are assigned a label of 0.

Next, we use the design features and the class labels to set up a supervised learning problem that can be solved using the decision tree-random forest method. The method is trained to differentiate designs that meet the target performance (with label 1) from those that do not (with label 0) based on the design feature values. Figure 2C shows a sample decision tree for the design problem of the Miura-ori unit cell. To determine the class of a data point, we send the data point into the tree from the top. Each time the data point encounters a branch node, a simple criterion is checked to determine if the point should go to the left branch or to the right branch. For example, a datapoint with $t_p = 1.1\ mm$ at the first node of the sample decision tree (shown on Fig. 2C) will be sent to the left branch because $t_p = 1.1\ mm < 2.1\ mm$. After a series of judgments, the data point will be sent to a leaf node, where no more branching occurs, and a class label is predicted. For example, the datapoint listed in column one of the Table in Fig. 2B will follow the gray arrow (marked as "Rule 1") downward and be judged as a Class 1 data. This means the machine learning algorithm *thinks* that this feature design ($t_c = 0.78\ mm$, $t_p = 1.1\ mm$, $W = 1.2\ mm$, $\gamma = 52°$) is most likely to produce a single unit Miura-ori with axial stiffness $k < 6000\ N/m$. The algorithm comes to this conclusion because the feature design values match the rule: $t_c < 0.8\ mm$, and $t_p < 2.1\ mm$. As such, each branch associated with Class 1 in the decision tree gives a design rule that would produce an origami design that meets the target performance. This highly interpretable structure of a tree method is useful for inverse design because the inverse relationship (i.e. $f^{-1}$) shows how to pick features based on the target performance. Moreover, we use randomly selected sub-datasets to train different decision trees to create more potential design rules. Training multiple decision trees forms an ensemble version of the tree method called a *random forest*. The structure of the trees, the splitting criteria, and the leaf node predictions are learned by the machine learning algorithm during the training process using a machine learning package called sklearn [36]. After training multiple decision trees, we gather all the design rules by tracing back through the tree branches (Fig. 2D). For example, Rule 1 and Rule 2 gathered in Fig. 2D are correlated to the two different branches in the sample tree (gray arrows in Fig. 2C).

Although these decision trees are automatically learned by the machine learning method, there are other manually specified variables that control how decision trees are computed. These user-specified variables are referred to as hyperparameters in machine learning and a technique called grid search is usually performed to select these hyperparameters. Different combinations of the hyperparameters are used to train the machine learning algorithms and the best combination is selected. Details on the considered hyperparameters and the grid search are provided in the supplementary materials section S3.3.

Now that we have collected a number of potential design rules, we need to select those that provide *better* performance for the inverse design problem. More specifically, a better design rule need to have a higher *precision* value and a higher *recall* value, where a high precision means that the rule is accurate, and a high recall means that the rule is representative (see the method section for details). In this work, we use the following routine to select design rules with better performance. The rules need to satisfy two thresholds and they are: (1) the precision is greater than 0.9, and (2) the number of data points satisfying the rule is greater than 10. Rules that do not satisfy these thresholds are eliminated from further consideration. Next, we rank the rules using the F-score function [37]:

$$F_\beta = (1 + \beta^2)\left(\frac{\text{precision} \cdot \text{ recall}}{\beta^2 \cdot \text{precision} + \text{recall}}\right).$$

This F-score function creates an average score from both the precision and the recall value, where the recall is seen as $\beta$ times more significant than the precision value. In this work, we select a value of $\beta = 0.2$ because the precision is more important for an inverse design problem, where the goal is to find "some designs" that meet the target performance not "all designs" that meet the target performance. Finally, we select the rules with the highest F-scores for our origami design and Fig. 2F shows the selected rule for this demonstration example. The shaded (darker) region of the box chart indicates the computed range of the four design variables and the design rule is $t_c < 0.78\ mm$, $t_p < 2.2\ mm$, $W \geq 1.9\ mm$, $\gamma \geq 67°$.

Although we managed to find a rule that performs well in the training dataset with a precision of 1.0, we need to further test the rule to verify that it is indeed a good rule. The testing is conducted by computing the precision of the rule using another testing dataset that is not used for training the machine learning method. This process is usually referred to as the *hold-out* testing in machine learning. Basically, the training data are homework problems for the machine learning algorithm and the testing data is the final exam. Details of the hold-out testing setup can be found in the supplementary materials section S3.3. In this demonstration example, a testing precision of 0.86 is obtained, which is reasonably good for an unbalanced dataset like the one for the single unit Miura cells, where the target data only consists of 7% of the total dataset.

Once good design rules are obtained and their quality is confirmed through testing, a designer can directly use these rules to design a suitable origami structure. Figure 2F presents one sample design that satisfies the rules and has the following features: $t_c = 0.70\ mm$, $t_p = 1.5\ mm$, $W = 2.5\ mm, \gamma = 70°$. This design has an axial stiffness $k = 5702\ N/m$, which indeed meets the target performance.

**Comparing Different Origami Patterns**

It is difficult to inverse design functional origami considering multiple origami patterns because *categorical features* are needed to represent and compare these different patterns. Categorical features including the type of origami pattern, the number of unit cells, and the topological design of each pattern, cannot be implemented directly into common continuous optimization-based inverse design methods [14, 15]. Capturing and comparing different patterns is essential because different origami produce intrinsically different motions and functional performances. To address this challenge, this section shows that the decision tree-random forest method can handle the complex interaction between categorical and continuous variables, which allows the method to compare and select between different origami patterns.

Here, we study a design problem for origami metasheets shown in Fig. 3A. These origami metasheets are cut out from a thin square plate with a footprint of $0.2\ m \times 0.2\ m$ and can be built from two distinctly different origami patterns including the standard Miura-ori pattern (Pattern 1) and the Tachi-Miura Polyhedron (TMP) [29] (Pattern 2). To represent these two patterns, we introduce an integer (binary) variable $p = \{1,2\}$. In addition, these origami metasheets can have different numbers of unit cells, represented as $m = \{24, 30, 36\}$ and $n = \{6, 9, 12\}$, in the two directions. In addition to these three categorical features, three continuous design features are also used in this problem, and they are the thickness of panels ($1.0\ mm < t_p < 6.0\ mm$), the thickness of creases ($0.5\ mm < t_c < 1.0\ mm$), and the width of creases ($1.0\ mm < W < 4.0\ mm$). An origami database of 2000 Miura origami samples and 2000 TMP origami samples is populated by randomly sampling values of the other design features. The origami simulation package SWOMPS is used to calculate the stiffness performance of the different origami systems (details in the supplementary materials section S2). In this example, we separately design the metasheet for two targets including the axial stiffness $k_a$ at 60% extension and the bending stiffness $k_b$ at 90% extension (design for multiple targets/objectives is discussed in the next section). As before, we define the extension of the origami ($Ext$) as the ratio between the folded length to the flat length. Four stiffness target zones are created for both the axial stiffness and the bending stiffness as indicated in Fig. 3B and 3C.

We first study the design for axial stiffness $k_a$ at 60% extension (see Fig. 3B). Assume that we want to design an origami metasheet to have $15000\ N/m < k_a < 30000\ N/m$ (Zone a1), we can label data that meet the target as 1 and the rest of the data as 0. After labeling, we can use the procedure in Fig. 2 to compute the design rules for this target. The design rule with the best performance is shown in the left column of Fig. 3B. We can then repeat the process for the other three target zones (Zone a2 to a4) by reusing the existing database. The computed rules for all targets are shown in Fig. 3B, and this series of rules tells us how the preferred design for an origami metasheet changes as the target for axial stiffness is increased. Interestingly, the machine learning method prefers changing the continuous variables to achieve the different axial stiffness targets without paying much attention to the categorical features used to represent different origami patterns. More specifically, the machine learning method suggests that controlling the thickness of creases ($t_c$) and the width of the creases ($W$) are more important than other parameters because tighter thresholds are used for these two features. Two sample metasheet designs for target Zones a1 and a3 are shown in Fig. 3D. The resulting performance of these designs indeed falls within the desired targets.

However, the design rules can be very different when we study the bending stiffness at 90% extension. Similarly, four design rules are computed for four different target zones as shown in Fig. 3C. Here, we set the targets to contain less data points in order to test out how well the proposed method performs when dealing with unbalanced databases, where the targets only contain around 5% of the total data. When we investigate the result of this series of design rules, we see that the machine learning method is paying more attention to the categorical features. As the target moves from one zone to another, the computed design rules change in a non-continuous manner because of the complex interactions between categorical variables and continuous variables. For example, when the target changes from Zone b2 to Zone b3 where we have a stiffer target, the machine learning method indicates that increasing the thickness of the panel is sufficient to meet the target. However, when we further increase bending stiffness requirement as we move from Zone b3 to Zone b4, the method suggests that it is better to change the categorical features (number of cells) and the continuous features simultaneously. A similar categorical jump is also observed when the target moves from Zone b1 to Zone b2. The proposed method can capture these complex interactions between the continuous features and categorical features, which cannot be done with optimization-based design methods.

## Multi-Objective Design for Multi-Physical Functional Origami

Handling multiple objectives is often necessary for designing functional origami structures because these systems exhibit multi-physical behaviors that need to be measured and compared using several different indices. Because such multi-objective problems are difficult to handle with standard optimization-based techniques, most existing functional origami systems were designed using trial-and-error approaches [4, 5, 6]. In this section, however, we demonstrate that the tree method can effectively handle multi-physical behaviors and can simultaneously consider multiple objectives. To this end, we present an example design for an electro-thermally actuated origami gripper where dynamics, power consumption, thermal behavior, and stiffness are all of interest (see Fig. 4).

In this example, one of three origami gripper patterns can be selected to achieve the target gripping motion (closing the gripping tip to less than a 1 mm gap). We assume that the gripper is actuated at the creases using an electro-thermal bi-layer system demonstrated in Fig. 4A and discussed in detail in [7]. This actuator contains two material layers with different coefficients of thermal expansion where one layer also serves as an electro-thermal heater (the top layer in this case). Joule heating causes differential expansion in the two layers, local curvature at the crease, and global folding of the origami patterns. Design features for the gripper include one categorical variable $p$ used to describe the pattern, and seven continuous variables including the length ($3.0\ mm < L_1 < 7.0\ mm$) and width ($2.0\ mm < L_2 < 4.0\ mm$) of the gripper arm, the location of the first hinge ($0.2 < Ra < 0.6$), the width of the actuator creases ($100\ um < W < 300\ um$), the thickness of the two layers in the actuator design ($0.15\ um < t_1 < 0.25\ um$ and $0.6\ um < t_2 < 1.0\ um$), and the thickness of the panels $10\ um < t_p < 30\ um$).

We use four indices to compare the multi-physical performance of the gripper, specifically (1) the fundamental frequency ($freq$) of the gripper, (2) input heating power ($Q$) needed to close the gripping arm, (3) maximum crease temperature ($T$) during the gripping motion, and (4) stiffness ($k$) of the gripper in resisting loads applied to pry it open. The origami simulation package SWOMPS is used to simulate and find these four performance indices (see supplementary materials S2 for details).

To demonstrate how interpretable machine leaning can tackle multiple multi-physical objectives, suppose we want to design an origami gripper to simultaneously match the following performance indices: $10\,Hz < freq < 40\,Hz$, heating power $Q < 0.2\,W$, and maximum temperature $T < 200°C$ (Target 1). We label all data points that satisfy the performance targets to be Class 1 and label the rest of the data as Class 0. Then, by computing the more representative decision rules for Class 1, we obtain design rules for a functional origami that will satisfy all three performance indices simultaneously. Figure 4B shows two design rules that have the highest F-score for this Target 1. Interestingly, both rules have high precision and are nearly identical except for small differences in the selection of $t_1$ and $t_p$.

We can use the same method to simultaneously design for all four performance indices (Target 2: frequency $10\,Hz < freq$, heating power $Q < 0.7$ W, and maximum temperature $T < 500°C$, and stiffness $0.002\,N/m < k$), with the result presented in Fig. 4C. If we compare the computed rules for Target 2 with those for Target 1, we can see that the machine learning method has picked another pattern after adding in the minimum prying stiffness requirement. While Pattern 3 was selected for Target 1, its horizontal creases cannot provide additional stiffness to resist prying, so Pattern 2 is now selected to achieve Target 2. This high interpretability of the tree methods helps users to better understand and reason about the desired behaviors of functional origami systems. Moreover, the machine learning method also shows how important each feature is for different design targets. For example, controlling the values of the gripping arm length $L_1$ and the location of the first creases (defined by $Ra$) is only important for Target 2 but not for Target 1.

In general, the design rules with the highest F-scores obtained from the machine learning method tend to be similar to each other. However, Fig. 3D shows an interesting result where the top two competing rules have relatively large differences between them. The large difference is because the two rules select different origami patterns, and the machine learning method thinks that both of these rules are appropriate for the desired design. Rule 1 of Target 3 selects Pattern 1 while Rule 2 selects Pattern 2. These results highlight that by computing multiple rules with high F-scores, it is possible to find *distinct design alternatives* that can all achieve the desired performance. Because Target 3 is 2-dimensional, we present the data points that fit the rules in both the training and testing datasets in Fig. 3E. The result shows that the extracted data points can trace the design boundary nicely and fill the design boundary with reasonable coverage.

The results from this section demonstrate the capability for the tree method to design functional origami systems where multiple objectives with multi-physical performance indices are of interest. Moreover, the presented examples show that this methodology can provide alternative design options when they are available.

**Design for Non-Geometrical Properties Together with Origami Shape Fitting**

Finally, we demonstrate how the proposed method can enable origami shape fitting algorithms to further consider non-geometrical properties of the origami so that a holistic design can be accomplished. So far, most research on origami inverse design focuses on geometric shape fitting (such as those in [14, 15, 17]). Usually, the shape fitting problem can be constructed as an optimization problem, where the error between the target geometry and the origami is minimized given certain constraints [14, 15]. However, these existing shape fitting studies cannot consider the non-geometrical properties that determine the functional performance of origami systems. Moreover, these shape fitting algorithms often leave tremendous flexibility for a designer to vary the origami pattern (e.g. number of panels used or maximum size of panels), without showing which combination may be better. Thus, this section will demonstrate how the proposed method can enable existing origami shape fitting algorithms to consider the interaction between the shape fitting and non-geometrical behaviors of the origami systems.

As a demonstration, we implement our method on top of an existing shape fitting approach introduced in [14], where an analytical solution was derived to build Miura-origami strips to fit arbitrary planar curves. Figure 5A shows how this shape fitting method can generate different origami strips to fit a target planar curve. In this method, the target curve is first separated into a specified number of segments defined by $m$. Then a planar origami strip geometry is generated by setting the offset length $l_o$ of the center node and the width of the strip $W_s$. Finally, the 3D origami is created by extruding the planar geometry to form the Miura geometry with an extrusion length $l_e$. Figure 5A shows the shape fitting results for three different curves. As can be seen, there is great flexibility in selecting these parameters for shape fitting and the selection can now depend on which combination gives a more desirable non-geometric performance. Suppose our target is to build an origami structure that can achieve a given stiffness performance while fitting a target shape, how should we select these shape fitting parameters and other design features of the origami? The proposed machine learning based method is able to answer questions like this.

Without loss of generality, we focus on designing a Miura-origami half-circle arch with a 2m radius. The origami arch database is generated by randomly picking shape fitting design features and other origami design features. The shape fitting features include the number of segments ($m = \{8, 12, 16, 24\}$), the offset length ($100\ mm < l_o < 300\ mm$), the strip width ($10\ mm < W_s < 40\ mm$), and the extrude dimension ($50\ mm < l_e < 250\ mm$). Other design features such as the thicknesses of panels ($1.0mm < t_p < 6.0\ mm$) and creases ($0.5\ mm < t_c < 1.0\ mm$) and the width of creases ($1.0\ mm < W < 4.0\ mm$) are also included because they affect the stiffness of the arch. Based on these features, we compute and record the responses of 3000 random origami designs using the SWOMPS simulation package to populate a database. The performance indices include the stiffness in X-direction ($k_x$) and Z-direction ($k_z$), the error of shape fitting ($e$), and whether the structure will snap or buckle under a 5 N load applied vertically ($S_z$) (see supplementary materials section S2 for details).

With the database established, we apply the proposed inverse design method to analyze the database. Because the decision tree-random forest method can handle a mixture of categorical and continuous variables, it can consider the integer variables used in shape fitting algorithm. Moreover, because the method can also tackle multi-objective problems, we can design for different combinations of shape-fitting errors and stiffness targets. Target 1 in Fig. 5C represents a target with stricter stiffness requirement but a less strict shape fitting objective while Target 2 has a more relaxed stiffness requirement but a stricter error objective. Both targets contain about 5% of the total data, so they are comparable in terms of overall design selectiveness. Figure 5C shows the computed decision rules for the two targets and indicates that both rules have reasonably high precision. This result demonstrates that the machine learning method can produce different design rules to accommodate the interactions between desired shape fitting error and mechanical performance. More importantly, the proposed machine learning method is not tied to specific origami patterns or shape fitting methods. Thus, the proposed methodology can be combined with other origami shape fitting approaches, such as those in [14, 15, 17], to enable a holistic inverse design of origami that considers both shape and non-geometrical function or performance.

Finally, we show that the proposed method can design origami systems with complex mechanical behaviors such as bistability and multi-stability [38, 39]. Target 3 in Fig. 5C shows the design rules for an origami arch to exhibit a snap-through behavior under a vertically applied 5N load. Unlike designing for a stiff arch with small fitting errors, the machine learning method shows that only the designs with more segments (larger $m$) in the strip can experience the snapping behavior. Moreover, it is necessary to have a low panel thickness and a low crease thickness so that the origami is more likely to snap. The testing precision of this design rule is high, indicating that the design rule is reliable and accurate.

**Discussion:**

In summary, this work establishes a novel inverse design method for functional origami structures using an interpretable machine learning method. To test the performance and versatility of the proposed method we built databases for four design scenarios, including: (1) Stiffness of a single unit Miura-ori, (2) Stiffness of origami metasheets, (3) Multi-physical performance of electro-thermal origami grippers, and (4) Stiffness and shape fitting of origami arches. These four databases can be reused for generating new designs or for testing the performance of other machine learning algorithms when applied to analyze origami related data. After populating the databases, the decision tree-random forest method was trained to identify origami patterns and design features in origami structures that achieve a specified target performance. Finally, origami design rules were computed by backtracking the splitting criteria associated with each tree branch and selecting the rules with the highest F-score. From a machine learning perspective, the proposed method uses the F-score, which is an averaged measure of precision and recall, to select better tree branches from the rest. Focusing on a few high performing branches in the decision tree-random forest method produces design rules with high interpretability for solving the inverse design function.

There are a number of benefits of using this data science and interpretable machine learning based approach for origami inverse design when compared to using existing optimization-based strategies. First, any generated database can be reused to compute new rules for different targets, or in other words to find appropriate designs for different scenarios. Second, the proposed method can simultaneously analyze the significance of design features for a given design target, which is not provided in optimization-based design methods. If a design feature is important for achieving a given performance target, a relatively tight threshold of that feature will be identified. Third, the proposed method demonstrates the complex interaction between continuous variables and categorical variables. Identifying these interactions is necessary for designs where comparing different origami patterns will naturally introduce categorical variables that cannot be captured with continuous optimization-based design methods. Fourth, we show that the proposed method can handle multi-objective design targets for active origami systems with multi-physical behaviors. Finally, we demonstrate that the proposed method can extend existing origami shape fitting algorithms to further design for non-geometrical performance of origami structures, which together enables a holistic framework for inverse design of functional origami. We envision that the proposed methodology can be used for designing active origami systems with superior performance for various applications in biomedical devices, soft robotics, metamaterials, deployable structures, and many more.

**Methods:**

**Origami simulation:** This work uses an open-access origami simulation package called SWOMPS [33]. This origami simulator uses a common simulation technique called the bar and hinge model to represent the geometry of origami systems. This simulator can explicitly model compliant origami creases (folds with distributed width) which makes it suitable for simulating the behaviors of practical origami structures [34]. In addition, the simulator package integrates a state-of-the-art multi-physics model to capture the electro-thermal actuation [35] important for active origami assemblages. The implementation codes for building origami databases are thoroughly discussed in the supplementary materials and are available on the GitHub page: https://github.com/zzhuyii/GenerateOrigamiDataSet. A more detailed introduction of the underlying origami simulation method and how it is applied in each design example can be found in the supplementary text sections S1 and S2.

**Machine learning method implementation:** This work uses an open-access package sklearn [36] to implement the decision tree-random forest machine learning method. More specifically, the sklearn package implements a classical decision tree method, where the trees are trained using suboptimal greedy approaches. More state-of-the-art tree methods (like the Generalized and Scalable Optimal Sparse Decision Trees [40] and the Optimal Classification Trees [41]) could be used in the future to further improve the performance. When training decision trees, an entropy-based criterion is used to identify the best splitting rules at branch nodes. Because the target class tends to contain only a small number of data (5% to 10%), the balanced class weight is used to tackle the imbalanced dataset. The results computed in the main article are accomplished using the following hyperparameters: the cost-complexity-pruning alpha value is 0.001, the maximum depth of trees is 20, and the number of training trees is 100. The details on the hold-out testing and hyperparameter selection can be found in the supplementary text section S3. All codes for the machine learning analysis and all datasets are discussed in the supplementary materials and can be found on GitHub: https://github.com/zzhuyii/TreeForOrigami.

**Selecting better design rules:** Common evaluation indices for machine learning, including the precision and the recall, are used to evaluate the performance of design rules. The precision is defined as the ratio between the number of accurate predictions of Class $t$ over the number of all predictions of Class $t$ [37]. In Fig. 2E, Rule 1 (blue dots) predicts 10 data points as Class 1, and 9 of them are correct so it has a precision of 0.9. The recall is defined as the accurate predictions of Class $t$ over the number of all data of Class $t$ [37]. In Fig. 2E, suppose we have a total number of 30 points in the target zone (within the blue decision boundary), then Rule 1 will have a recall of 0.3 (9/30).

**Acknowledgments:**

The authors want to thank the helpful discussion with Ziyang (Tony) Tang at Purdue University.

**Funding:**

National Science Foundation Grant #2054148

**Author contributions:**

Conceptualization: YZ

Methodology: YZ

Investigation: YZ

Visualization: YZ

Supervision: ETF

Writing—original draft: YZ

Writing—review & editing: YZ, ETF

**Competing interests:**

The authors declare they have no competing interests.

**Data and materials availability:**

All data and code used for the analyses are available in the main text or the supplementary materials.

## Figures and Tables:

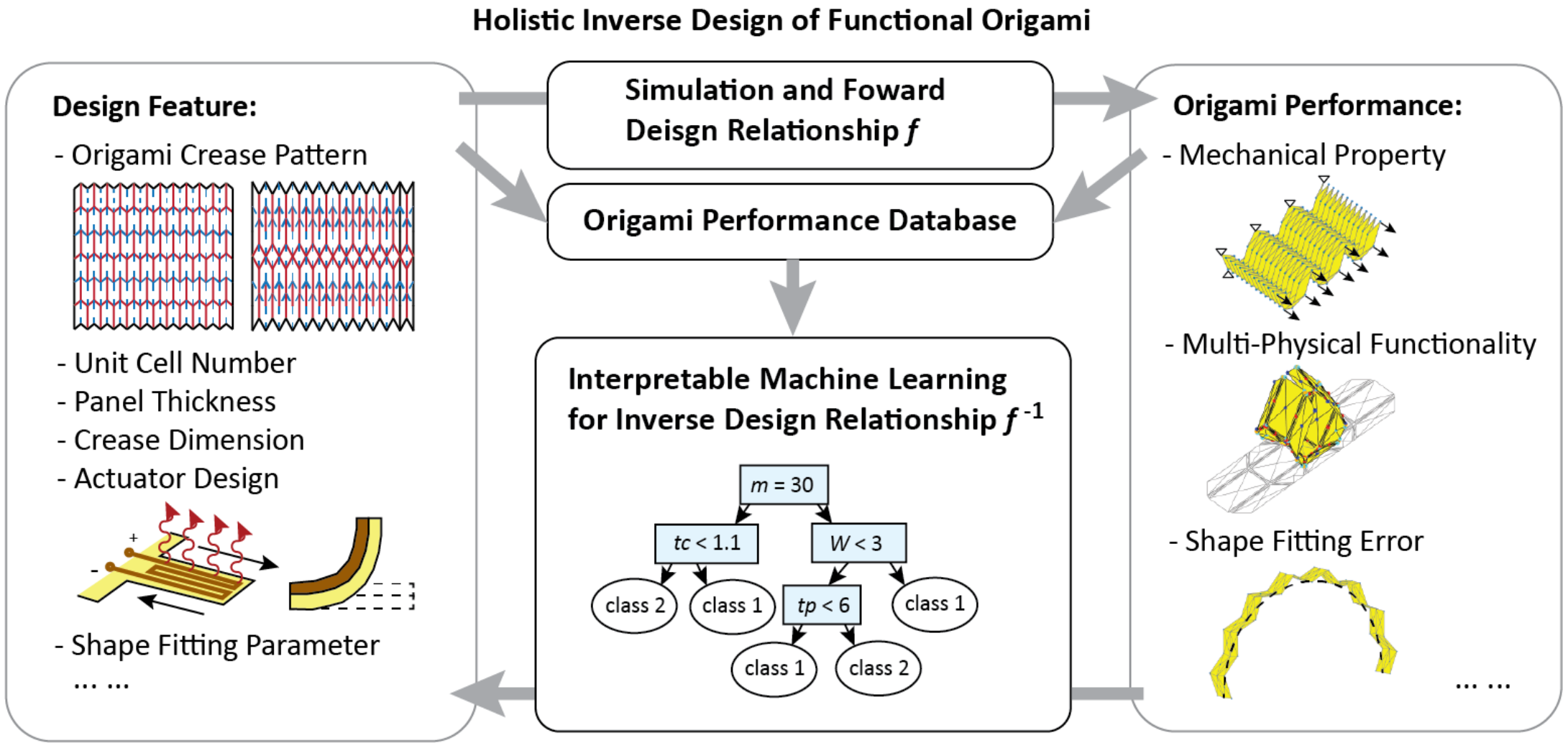

**Fig. 1. Interpretable machine learning for inverse design of origami.** The relationship between origami design features (left) and performance of origami systems (right) can be thought of as a black-box nonlinear function $f$. This work shows that it is possible to train an interpretable machine learning method (a decision tree-random forest method at the bottom) to uncover the underlying structure of this black-box nonlinear function $f$, so that we can build human understandable design rules to solve the inverse design problem (solve for $f^{-1}$).

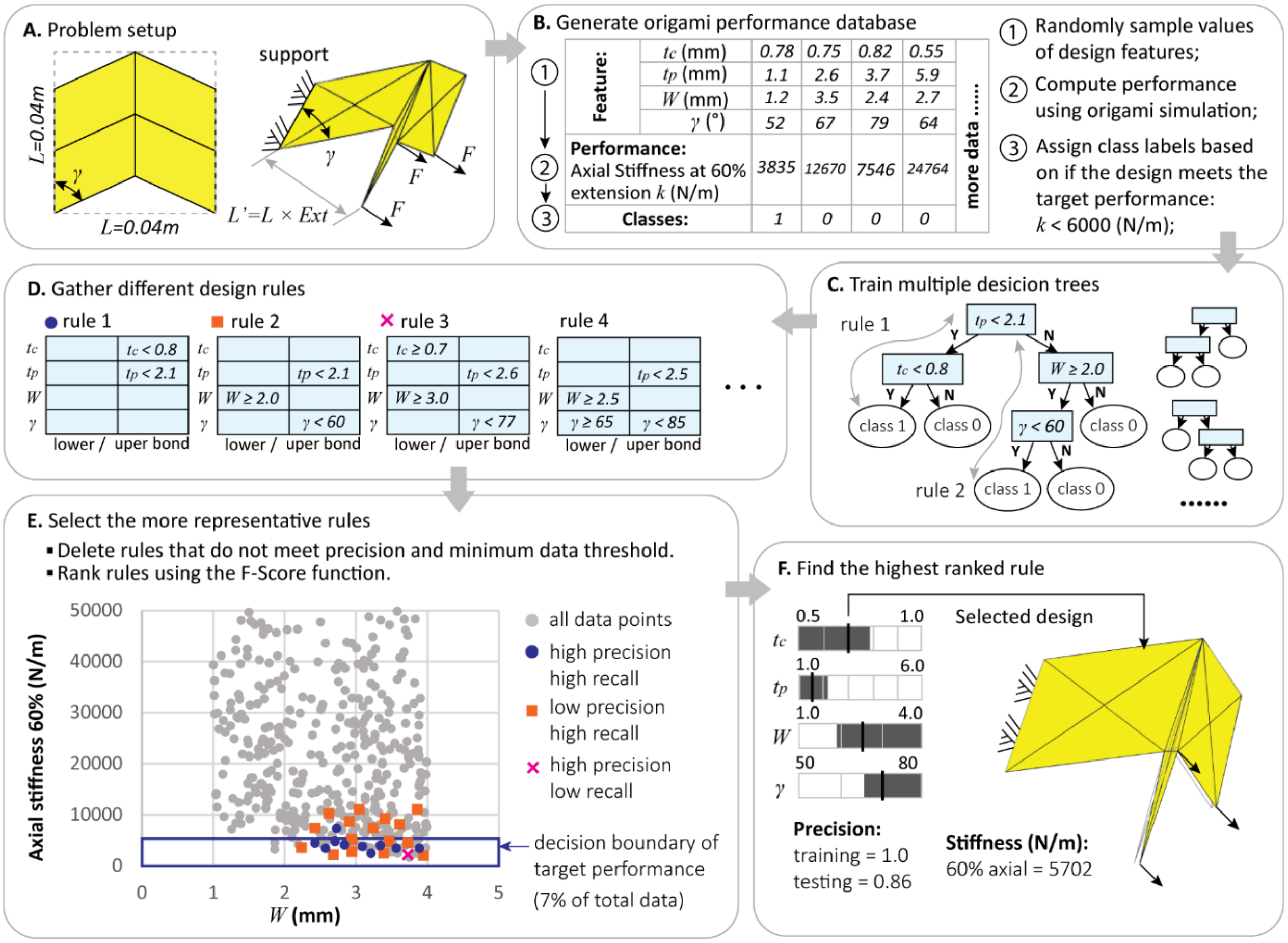

**Fig. 2. Computing interpretable design rules for origami assemblages. A**. The problem setup for the single unit Miura-ori cell. The axial stiffness is computed for the folded origami with supports on the left and axial loads on the right. **B.** A database of Miura-ori design features and resulting axial stiffness performance is populated using a physics-based origami simulator. Data points are labeled based on whether they meet the target performance or not (axial stiffness $k <$ 6000N/m). **C.** A number of decision trees are trained to classify the database. **D.** Design rules are gathered by collecting the splitting criteria in each branch of the decision tree (following the gray lines in the sub-figure (C)). **E.** Representative rules are selected based on the precision, the recall, and the number of data points satisfying the rules. In this case, the rule 1 (blue dot) is better than the rule 2 (orange box) and the rule 3 (pink cross) because it has high precision and recall. **F.** A final design rule with the highest F-score is picked. The box chart contains the full range for all features, while the shaded regions indicate the final design rule for the target performance. A Miura-ori design that follows the final rule is shown (indicated with dark lines in the box chart).

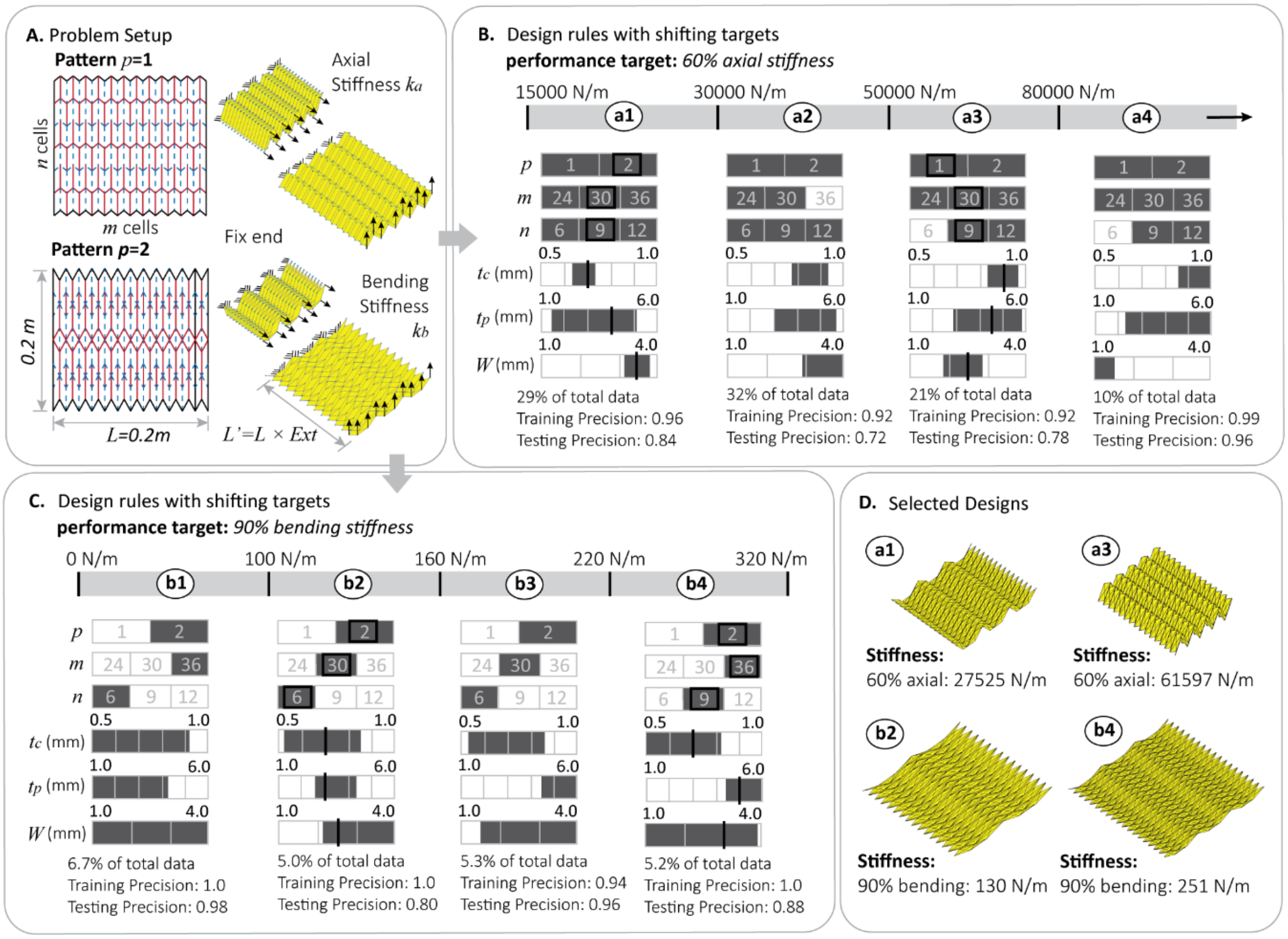

**Fig. 3. Comparing different origami patterns using the decision tree-random forest method.** **A**. Problem setup for the design of origami metasheets. Two origami patterns are compared for their performance as a metasheet. **B**. Design rules for different axial stiffness targets when the metasheet is at a 60% extension. **C.** Design rules for the bending stiffness at a 90% extension. **D.** Sample designs for the four selected zones in the axial and bending stiffness design and the performance of each sample.

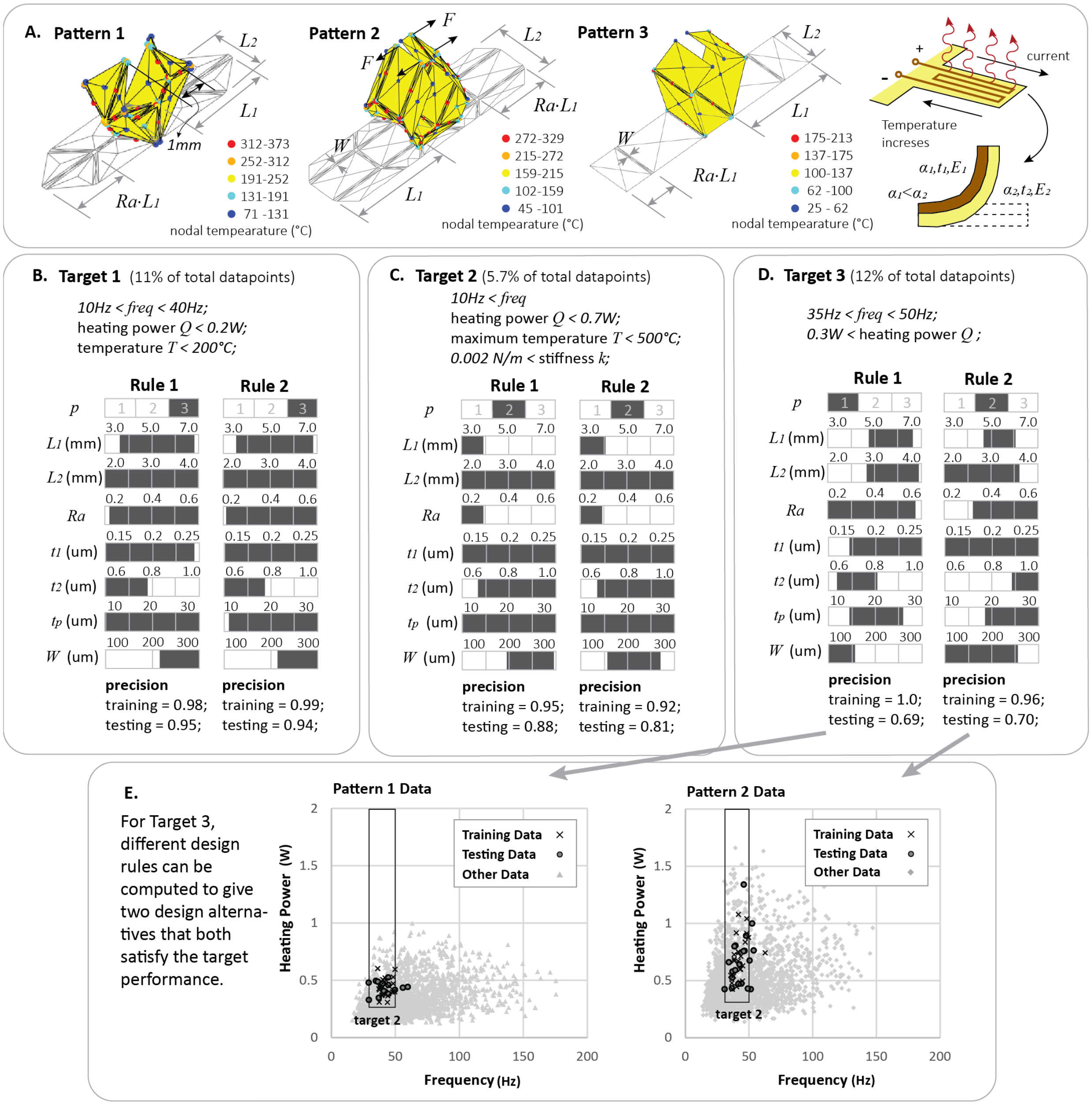

**Fig. 4. Design of an origami gripper with multi-physics and multiple objectives. A.** Problem setup for building active electro-thermal origami grippers with three different patterns. **B-D.** The top two design rules of designing the grippers for three different multi-objective targets (Targets 1 to 3). **E.** The top two design rules for Target 3 give distinct design alternatives where a different origami pattern is selected, yet both data point distributions meet the given target.

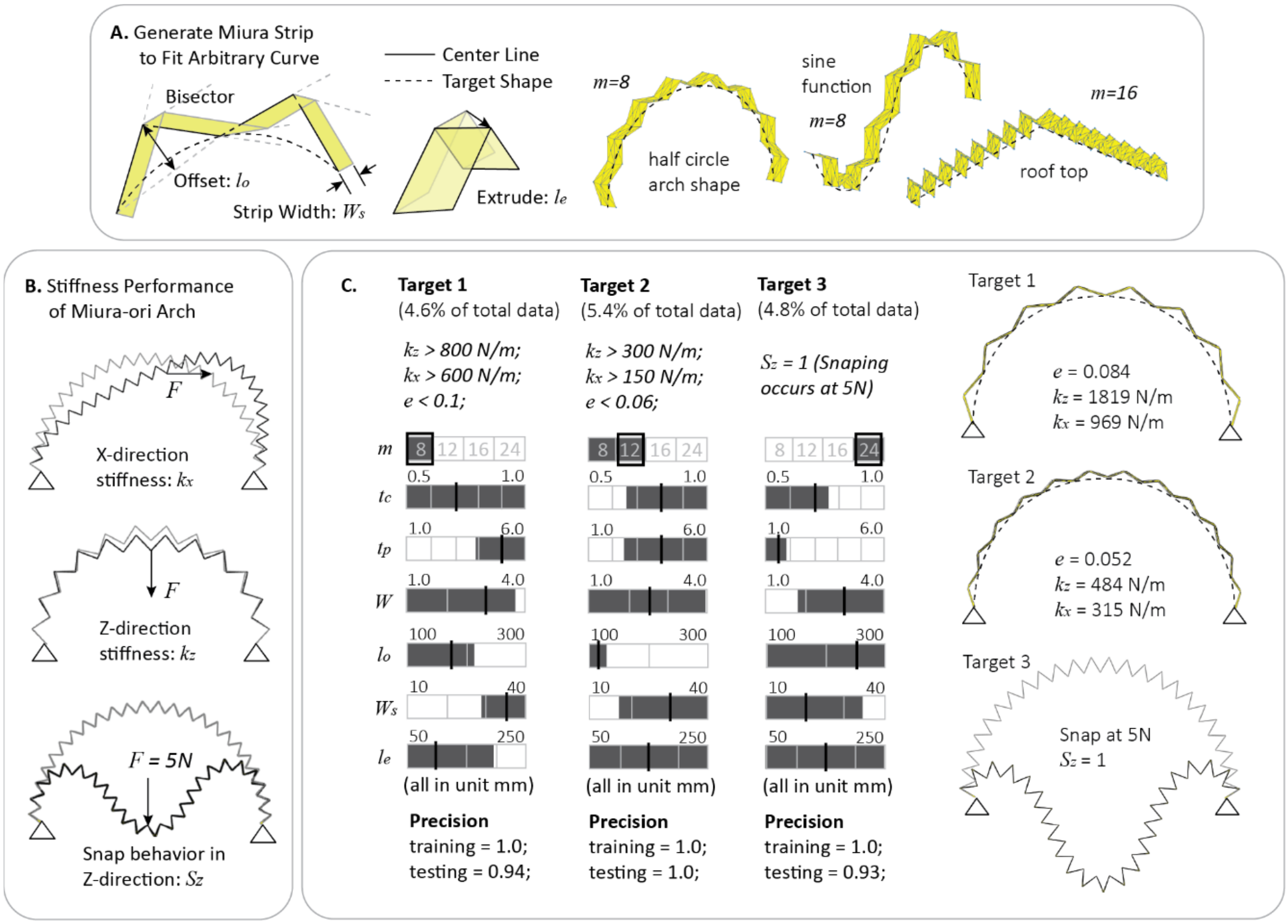

**Fig. 5. Integrating the machine learning method with shape fitting for holistic inverse design of origami. A.** Definitions for a modified Miura-ori design that can fit arbitrary curved shapes. Geometrical design features for the modified Miura include the number of units $m$, the offset length $l_o$, the width of the strip $W_s$, and the extrusion length $l_e$. **B.** Four performance indices are studied including the error of fitting $e$, the stiffness in X and Z directions $k_x$ and $k_z$, and a binary variable $S_z$ that indicates if the structure will snap if loaded with a 5N load in the Z direction. **C.** Computed rules for three different targets and corresponding sample designs.

# Supplementary Materials for

## Harnessing Interpretable Machine learning for Holistic Inverse Design of Origami

Yi Zhu*, Evgueni T. Filipov

*Corresponding author. Email: yizhucee@umich.edu

**This PDF file includes:**

Supplementary Text
Figs. S1 to S8
References (1 to 12)

**Other Supplementary Materials for this manuscript include the following:**

Data S1 to S4
Codes S1 to S2

## Supplementary Text

### S1. Introduction of Bar and Hinge Models

This section describes the bar and hinge model used to generate the origami performance database for the manuscript. The bar and hinge model is a widely used reduced-order model for simulating the mechanical behaviors of origami systems [1, 2], and it contains two major elements: the bar element and the rotational spring (hinge) element. The bar elements are extensional 3D springs that can capture in-plane behaviors of origami such as the stretching and shearing deformations (see Fig. S1B). The rotational spring elements are rotating hinges with stiffness that can capture the out-of-plane behaviors of origami including the panel bending and crease folding deformations (see Fig. S1B).

More specifically, this work uses the origami simulation package SWOMPS published in [3] where the bar and hinge model is enhanced to capture the multi-physical behaviors important for active origami systems. The SWOMPS package implements both the standard bar and hinge formulation where the folding creases are simplified as 1D rotational hinges and the compliant crease bar and hinge formulation where the compliant crease regions are captured with distributed widths [4] (see Fig. S1A). The package also supports the simulation of multi-physics-based electro-thermal crease actuation [5].

A potential-based static simulation is implemented in this simulation package. The nonlinear folding behavior of active origami systems can be traced using different nonlinear PDE solvers. In general, the internal potential of the active origami can be expressed as:

$$U_{ori}(\boldsymbol{x}, Q) = U_{bar}(\boldsymbol{x}) + U_{spr}(\boldsymbol{x}, Q),$$

where $\boldsymbol{x}$ represents the nodal coordinates of the origami system and the terms $U_{bar}, U_{spr}$ are the potential energies associated with the bar stretching and spring rotation respectively. The applied heating power $Q$ can affect the stress-free configurations of the rotational springs and thus, change the potential of springs. Linear elastic bar elements and rotational spring elements are used in this work so the potential of bar element is:

$$U_{bar} = \frac{1EA}{2L_0}(L(\boldsymbol{x}) - L_0)^2,$$

where $L$ is the bar length, $L_0$ is the original length, $E$ is the Young's modulus, and $A$ is the area of the bar element. Similarly, the potential of the rotational spring is:

$$U_{spr} = \frac{1}{2}k_{spr}\big(\theta(\boldsymbol{x}) - \theta_0(Q)\big)^2,$$

where $k_{spr}$ is the spring stiffness, $\theta$ is the rotational angle, and $\theta_0$ is the stress-free rotational angle which is a function of the applied heating $Q$. Detailed derivations of the stiffness parameters (e.g. $A$, $k_{spr}$) can be found in [1, 4].

Finally, Fig. S1C shows how the bar and hinge model can capture heat transfer within electro-thermally actuated origami structures. Because active origami structures tend to be small and operate at a low temperature, there is limited convection and radiation within the system. Therefore, we can capture the heat transfer within the origami system and heat loss from the origami structure to the ambient environment using simplified conduction models. The heat transfer within origami is captured using linear triangular thermal conduction elements and the heat transfer between the origami and the surrounding environment is simulated as a simplified 1D heat conduction problem (see Fig. S1C). Reference [5] provides further details on the formulation, verification, and validation of this simplified simulation method.

Solving for the equilibrium of the origami under applied loads or applied heating power $Q$ can be achieved by finding the local minimum (extremum) of the total potential energy:

$$\frac{\partial U_{ori}(Q)}{\partial \boldsymbol{x}} = F_{ext}.$$

A number of nonlinear PDE solution methods can be used to find the equilibrium configuration associated with a given loading or applied heating [5, 6]. The SWOMPS package provides the Newton-Raphson Method, Displacement Control Method, and the Modified Generalized Displacement Control Method for solving the equilibrium. In this work, one of these three methods is used to solve for the equilibrium of the origami depending on the type of the problem.

Using the SWOMPS package for generating the performance database is beneficial because of the following reasons. First, this package can capture the electro-thermally actuated folding creases so that we can study the power consumption and maximum temperature of active origami structures. Second, the SWOMPS package is computationally efficient, so it is possible to populate an origami performance database rapidly. Finally, this package is verified against experiments, so the populated database is representative of realistic origami behaviors. The bar and hinge origami simulation codes and the computed origami performance databases can be found on GitHub or in the supplementary material of this manuscript.

**S2. Generating Origami Performance Database**

This supplementary section describes how we populate the three more complex origami performance databases demonstrated in the main text. The execution codes for all three examples are published as supplementary materials and can be found on GitHub. Links to the GitHub website are provided in the final section of this supplementary text.

Figure S2 shows how we built the origami canopy database. Two different origami patterns are used and they are represented using a categorical feature $p = \{1,2\}$, meaning patterns 1 and 2. We also consider the number of cells in the x and y directions ($m = \{24,30,36\}$ and $n = \{6,9,12\}$, respectively) as categorical design features. Other design features are continuous and include the thickness of panels ($t_p$) and creases ($t_c$) and the width of the creases ($W$). We assume the material used to fabricate the origami has a Young's modulus of 2 GPa, which represents common polymeric materials. Furthermore, we use a standard bar and hinge model where the crease folding stiffness is represented using a single rotational spring. This spring stiffness value is calculated using a pseudo-rigid-body model (see Fig. S2C). The calculation of other stiffness parameters follows the method introduced in [4]. When solving for the global stiffness of the canopy structure, the left end of the origami are fixed in space (no translation in xyz directions) while the right end are loaded with uniformly distributed forces. We apply a total of 3 N force and use a Newton-Raphson loading scheme to compute the resulting displacement. The global stiffness is calculated as the ratio between the applied load and the nodal deformation. Two thousand (2000) samples with randomly generated feature values are calculated for each pattern, and the database contains both the bending and axial stiffnesses at 30%, 60%, and 90% extension. The extension is measured as the ratio between the folded length of the pattern $L'$ and the original length $L$ where the origami is a flat and unfolded sheet.

Figure S3 shows how we build the database for active origami grippers. Figure S3A shows the geometrical definitions of the three origami patterns used to build the grippers, where the red solid lines are valley folds and the blue dotted lines are mountain folds. A categorical feature $p = \{1,2,3\}$ is used to represent these three different patterns. Other design features are continuous and include the length ($L_1$) and width ($L_2$) of the gripper arm, the location of the first hinge from the base (measured as a ratio $Ra$ of the outer arm compared to the total arm length), the thickness of the two layers in the actuator design ($t_1$ and $t_2$), and the width of the actuator creases ($W$). All creases in these patterns are assumed to be bi-layer electro-thermal actuators similar to those in [7]. We assume that both the active layer and the passive layer of these actuator creases have a Young's modulus of $E_1 = E_2 = 2$ GPa and the difference between the thermal expansion coefficient is $\Delta\alpha = 50\times10^{-6}$. We also assume that the origami system has a density of 1200 kg/m$^3$, which is representative of common polymeric materials. The fundamental frequency of the system is computed using a particle bar and hinge formulation, where the mass of the structure is lumped at the nodes of the origami. In this example, a compliant crease bar and hinge formulation is used to represent the distributed geometry of the active creases, where the folding of an active crease is represented using three lines of rotational springs [4]. Electro-thermal heating is applied uniformly to all creases to trigger the actuation. We solve the applied heating power $Q$ needed to close the gripping arm by incrementally increasing the applied power with a step of 0.1 W/m (applied heating per length of creases). This electro-thermal folding motion is captured using a simplified origami simulation model introduced in [5]. After the gripper is closed with the actuator creases, a Newton-Raphson loading scheme is used to study the stiffness of the gripping arm. A small force ($1\times10^{-8}$N) is applied outward at the tip to calculate the stiffness of the gripper for resisting forces prying it open. Two thousand (2000) sample grippers for each pattern are simulated to create the database.

Finally, we present the details on how we build the origami arch database (Fig. S4). To fit an origami strip to an arbitrary curve, the target curve is first separated into different segments determined by the number of segments $m = \{8,12,16,24\}$. Then, the centerline of the Miura origami geometry is determined using the offset distance $l_o$, where the center node is moved towards one direction of the curve from the center point by $l_o$. After determining the centerline, the planar Miura geometry is determined using the bisector angles and the width of the strip $W_s$. Finally, the 3D Miura origami is generated by extruding the 2D planar geometry using the length $l_e$. Further details on this shape fitting method can be found in [8]. In this origami arch database, the target curve is a half-circle with a radius of 2m. After the Miura half-circle arch is generated, its mechanical performance can be calculated using the SWOMPS simulation package. The support of the arch is set by fixing the three nodes at each end of the strip in 3D space. We assume the material has a Young's modulus of 2 GPa and a Poisson's ratio of 0.3. A concentrated force of 5N is applied at the middle of the arch as shown in Fig. S4B. The deformed configuration is solved using a Newtown-Raphson solver, and the stiffness is calculated as the ratio between the applied load and the nodal displacement (as secant stiffness). The snapping behavior is determined by judging if value of the nodal displacement has exceeded 1m. Three thousand (3000) samples are calculated to build this database.

## S3. Compute Interpretable Design Rules with Decision Tree-Random Forest Method

In this supplementary section, we will introduce the decision tree-random forest method, which is an inherently interpretable machine learning method. First, we give a brief introduction on the formulation and the training process of the tree method. Next, we will talk about the hyperparameter selection and introduce one technique called one-hot-encoder to integrate categorical features into the sklearn implementation package for tree methods.

### S3.1 Introduction to Decision Trees and Random Forests

Figure S5A shows how a decision tree classifier works. There are two types of nodes within a decision tree and they are branch nodes and leaf nodes. Consider a classification problem where the decision tree is used to differentiate two classes. To predict the class of a data point, the data point can be sent into the decision tree from its root branch node (the branch node at the top). The datapoint will then either flow to the left or to the right depending on whether it meets the criterion associated with the branch node or not. Finally, after a number of judgements, the datapoint will arrive at a leaf node where it will be predicted to be either class 1 or class 2 (for this binary classification problem). As shown in Fig S5A, the formulation of a decision tree is highly interpretable and easily understood by humans. This is one major advantage of using interpretable machine learning methods over traditional "black-box" machine learning methods like neural networks [9].

Figure S5B shows how a decision tree classifier can be trained. The classical method for building decision trees is a sub-optimal greedy method. Instead of building the entire tree all-at-once with optimized performance, the tree is built in a node-by-node manner. Suppose we start with a training data set, which is a mixture of data classified either as class 1 or 2. Then, the decision rule of the root node is designed such that after the separation, the two new datasets are "more pure" than the original combined input dataset. The purity of the dataset can be measured using the entropy defined as:

$$\sum_{c=1}^{C} -\frac{n_c}{n} log\frac{n_c}{n},$$

where $c$ is the class of data, $n$ is the number of data, and $n_c$ is the number of class $c$ data. For a two-class classification problem, we have $c = \{1,2\}$. Based on this definition, the purer the data, the lower the entropy value will be. Other similar criteria such as the Gini index can also be used in addition to the entropy. When selecting the decision rule of a branch node, different potential rules are enumerated to find the one that can produce the lowest entropy after the separation.

After producing the two new sub datasets, we can then determine if we should keep splitting the new dataset or not (creating a new branch node). If the sub datasets are still not pure enough and the tree has not yet reached the specified maximum depth, the algorithm will redo the process and obtain a new splitting rule. However, if the sub datasets are pure or if the tree has reached the maximum depth, the algorithm will stop and generate the leaf nodes. The prediction label of a leaf node is determined by the most commonly occurring class of the training data separated into that leaf node.

Because the decision trees are built node-by-node, we can only obtain sub-optimal trees with this approach. There are also methods to create optimal decision trees such as the OCT method demonstrated in [10], but this formulation requires solving mixed-integer-programs, which is much more time consuming than using a greedy approach. Therefore, this work uses the standard greedy approach to generate the decision tree.

In practice, it is more common to use the ensembled version of the decision tree method, the random forest method, for better performance. The formulation of a random forest is straightforward, and it simply means generating different decision trees and taking the most common class prediction result out of all trees. One way of obtaining different decision trees is to use random subsets of the training data for the training process (see Fig. S5D). This work uses the random forest method to create drastically different tree branches for better inverse design rules.

### S3.2 Techniques for Integrating Categorical Features

In this work, the decision tree method is implemented using the widely used sklearn package [11]. Although the decision tree method can consider categorical data theoretically [12], conventional implementations of the decision tree (such as the one in the sklearn package) may not directly support the use of categorical data. In the default setting of the sklearn package, all variables are treated as continuous numerical values. However, there is a technique called one-hot encoder that can be used to resolve this problem.

Considering that we have a categorical feature called "type of origami pattern" $p$ and the values for this feature include $\{1,2,3\}$. Without using the one-hot encoder, the sklearn package will treat $m$ as a continuous numerical variable, which means that the machine learning algorithm can potentially create a rule in the form of "pattern $p > 0.5$". This is inaccurate because $p = 2$ and $p = 3$ are categorical features that cannot occur at the same time. To fix this issue, we can use the one-hot encoder to convert the integer input data (see Fig. S5C). The categorical variable $p$ ($p = \{1,2,3\}$) is converted to three different binary variables as "$p = 1$", "$p = 2$", and "$p = 3$", where each data point can take values of 1 for true or 0 for false. With this formulation, the sklearn package can treat the variable $p$ as categorical through using the encoded binary true/false variables. Using the one-hot encoder introduces a large number of sparse input features that can potentially make the algorithm computationally inefficient. In such scenarios, it would be more effective to select a dedicated implementation package of the tree method that can directly support categorical features.

### S3.3 Hold Out Test and Hyperparameter Selection

Here, we introduce the hold out testing setup and hyperparameter selection process that are needed to use the decision tree method. Figure S5D gives an illustration of the hold out testing setup. The idea of doing a hold out testing is to check the performance of our trained machine learning algorithm. What we do is to 'hold out' a number of data from the full database while training the machine learning algorithms and only use these data when we are testing the performance of the machine learning. It is similar to creating two sets of questions: one set for training the students, and one set for testing how well the students learned from working out the questions in the training set. In this work, the complete dataset is randomly split into two sets: one with 60% of the data for training the machine learning method, and the other one with 40% of data for testing the performance.

Hyperparameters refer to the variables in the machine learning algorithm that are explicitly specified by the users and are not learned from the training data. Usually, a grid search is performed to find the optimal hyperparameter values for the machine learning algorithm. In this work, we consider five hyperparameter variables and they are (1) the maximum depth of the trees, (2) the number of tree learners in the forest, (3) the cost-complexity pruning alpha value. (4) the splitting criterion, and (5) the subset ratio for training. The precision of the selected final design rule is used to assess the choice of each hyperparameter. Five runs of different hold out tests are used, and the mean precision is recorded. Because the depths of trees and the number of tree trainers are the two major hyperparameters of a decision tree-random forest method, we mainly focus on the effects of these two values while only briefly highlighting the others. In addition to the grid search results, we also trained two other machine learning algorithms for comparison, namely the kNN method, and Gradient Boosting method. The kNN method is usually used as a baseline method in machine learning because it has a simple formulation. In general, we expect the precision obtained from the target method be superior to that obtained using the kNN method. The Gradient Boosting method is a “black box” machine learning method that tends to perform well on many datasets. If the target method can achieve a precision that is close to that from the Gradient Boosting, the target method can be seen as competitive.

We perform three rounds of grid search for the origami canopy (Fig. S6), the origami gripper (Fig. S7), and the shape fitting origami arch (Fig. S8) databases. First, we give a brief discussion on the influence of the maximum depth of trees and the number of trees in the forest. We can see from Fig. S6 to S8 that there is a saturation effect in the maximum depth of trees. The precision improves when increasing the maximum depth but stops once the maximum depth reaches certain values. This saturation is expected because if tree-based learners are too deep, the leaf nodes will just have one or two data points, which is seen as “pure” by the algorithm. In this case, further increasing the depth of trees will not change the tree structure. Moreover, branches like these are not representative and cannot be used to compute reliable design rules. Therefore, it is recommended to set the depth of trees so that we barely reach that saturation point. In this work, we select a depth of 20. For the number of tree trainers in the forest, we found that a relatively large number of tree trainer is helpful for improving the precision. However, when the value is too large, the precision can also saturate. This saturation happens because we may be close to “enumerating all possible” tree branch formulations when using a large number of tree trainers. In this case, adding tree trainers may not generate new branches we want. Thus, a value of 100 is picked for the number of trainers in this work.

In Fig. S6, with the grid search result for the origami canopy dataset, we further study the effects of the cost-complexity-pruning alpha (ccp-alpha) value. The cost-complexity-pruning is another step in training decision tree methods where the machine learning method will automatically remove some branch nodes from a tree to avoid over-fitting. In general, the larger the alpha value is, the more likely the algorithm will remove nodes. As demonstrated in Fig. S6, the pruning alpha value does not have a significant influence on the precision of our prediction, and increasing the alpha value does not improve the precision for all cases. Therefore, this work uses a ccp-alpha value of 0.0001 in all the training performed.

Next, the grid search of the origami gripper database is used to study the splitting criterion. The sklearn package provides two different splitting criteria and they are the Gini index and the entropy index. The Gini index is similar to the entropy but is computed differently [11]. From Fig. S7 we see that entropy tends to provide better performance than the Gini index. Therefore, we use entropy as the splitting criterion in our work.

With the origami arch database, we further consider the effects of the subset ratio. To train an effective random forest algorithm, different decision trees need to be built [12]. One way of achieving this goal is to use a randomly generated subset of the training data to train the decision trees instead of using the entire training database. The subset ratio is the ratio between the size of this sub dataset when compared to the entire training database. The results of the hyperparameter grid search are shown in Fig. S8. We can see that a smaller subset ratio can improve the performance of the machine learning algorithm, because a smaller subset ratio can encourage the algorithm to generate drastically different branches for computing the better decision rules. Thus, we select a subset ratio of 0.5 in this work. We did not explore the potential of having an even smaller subset ratio value because that will make the training set too small.

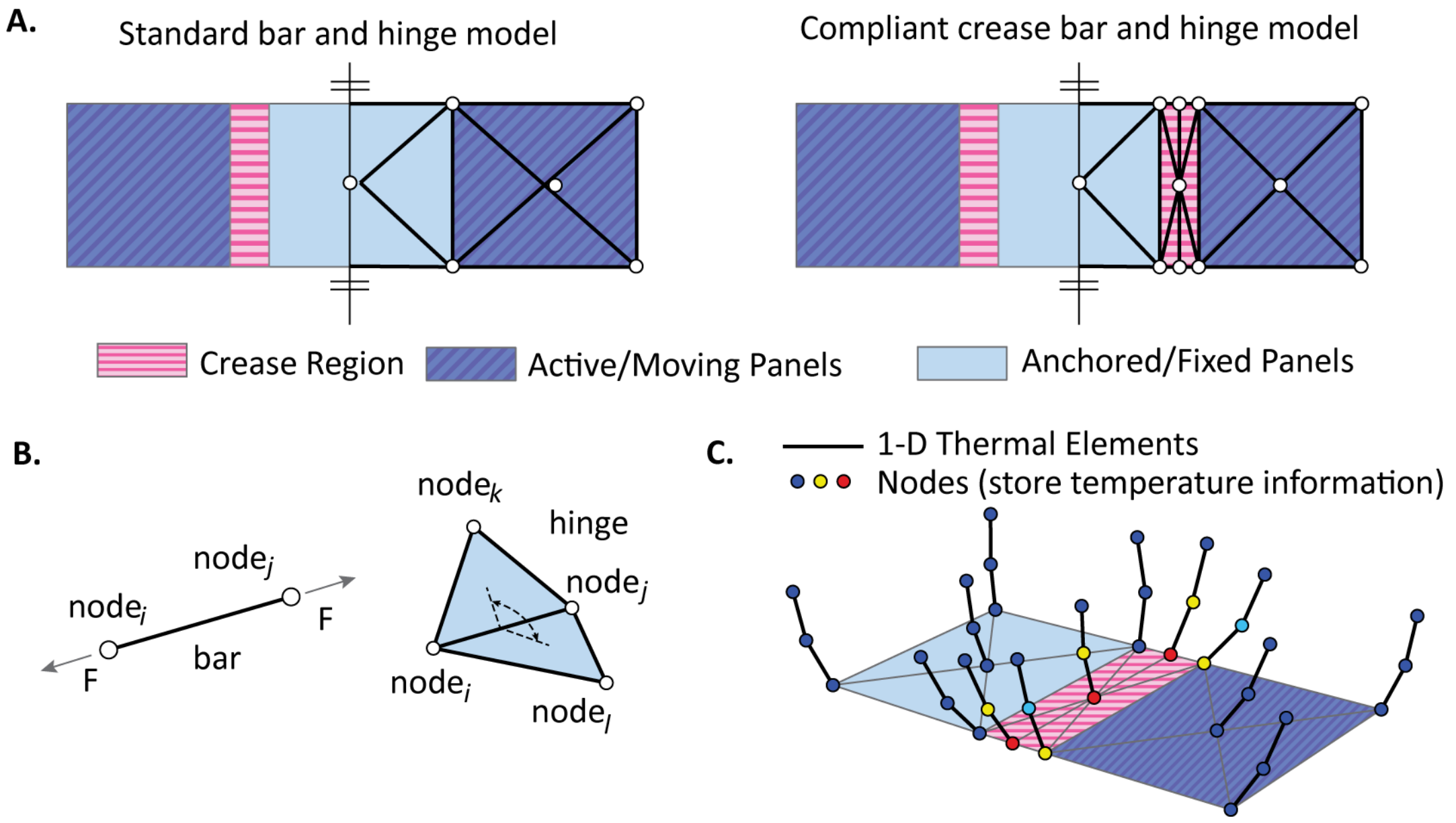

**Fig. S1. Bar and hinge models for simulating multi-physical behaviors of origami systems. A.** In this work, we use both the standard bar and hinge model where a folding crease is simplified to a 1D rotational element, and the compliant crease formulation that simulates the distributed crease region. **B.** The bar and hinge model uses bar elements to capture in-plane behaviors and rotational springs (or hinges) to capture the out-of-plane behaviors. **C.** A reduced ordered heat transfer model is used to capture the thermal distribution and heat-driven actuation of active origami systems.

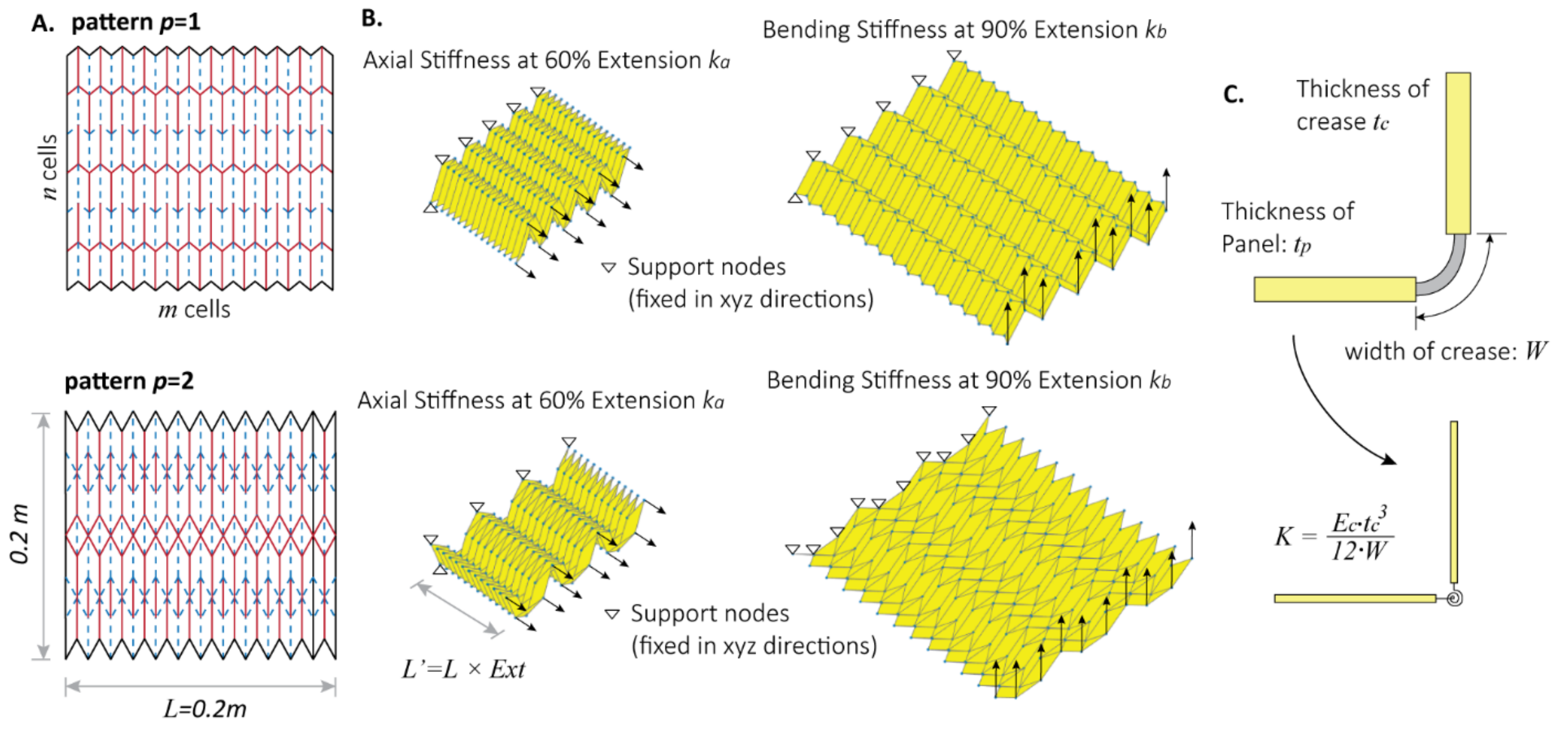

**Fig. S2. Details on building the canopy database. A.** The geometry of the Miura origami pattern ($p$=1) and the Tachi-Miura-Polyhedron (TMP) origami pattern ($p$=2). Both canopies are cut from 0.2m square sheets with the same thickness and material properties. **B.** The simulation setup for building the database. The nodes at the left end are fixed in 3D space while a load is applied on the right end. The global stiffness properties are calculated at different extension ratios (*Ext*) of the origami. **C.** A pseudo-rigid-body model is used to calculate the rotational spring stiffness of the creases in the origami canopies.

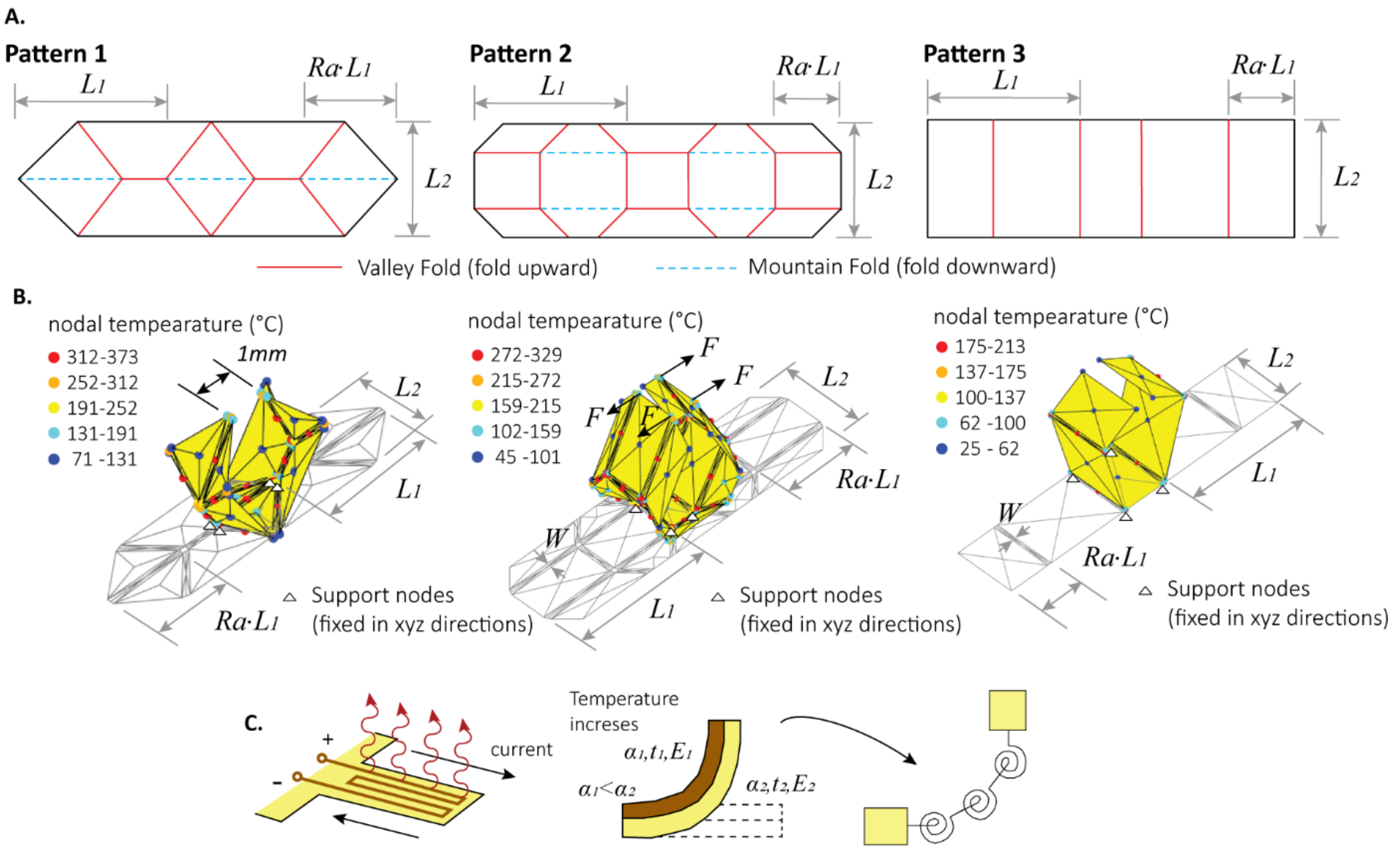

**Fig. S3. Details on building the active origami gripper database. A.** Geometrical definition of the three origami patterns used to build active grippers. **B.** Folding simulation of the three grippers. Four nodes at the base of each pattern are fixed in 3D space to serve as the supports. An electro-thermal loading is applied to close the gripper arms to be less than 1mm apart. Then, forces are applied onto the gripper tips to pry it open and the stiffness is recorded. **C.** A compliant crease bar and hinge model is used to model the electro-thermally driven bi-layer actuators at the active crease regions. The distributed curvature of these active creases is represented using three lines of rotational springs.

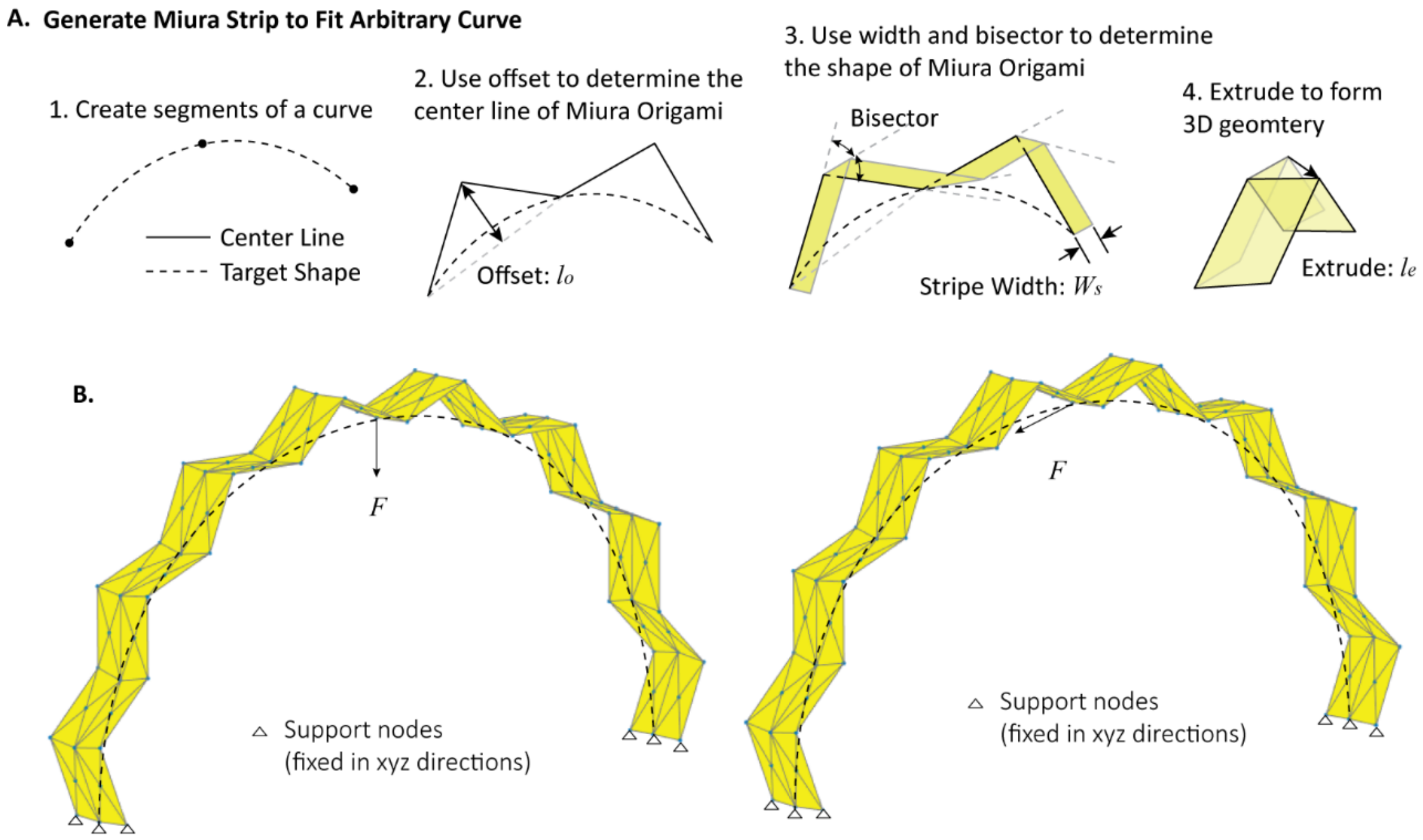

**Fig. S4. Details on building the database for origami shape fitting. A.** The four-step method to generate the Miura strip for shape fitting. **B.** Loading simulation of the origami arch. A concentrated force is applied at the center of the arch and the stiffness is calculated. The three nodes at each end of the arch are fixed in 3D space to serve as supports.

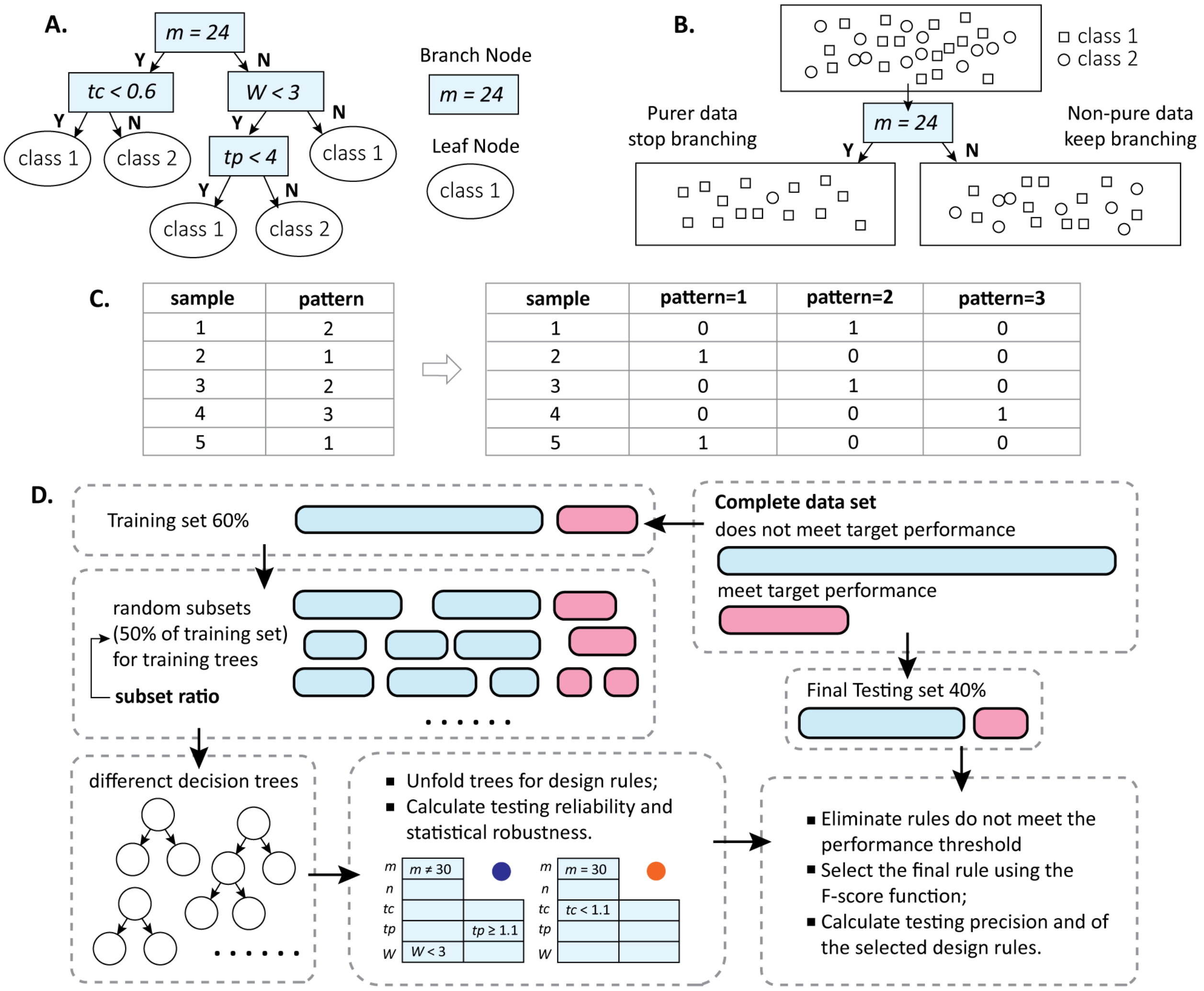

**Fig. S5. The formulation of a decision tree method. A.** A sample decision tree for a binary classification problem. **B.** The decision rule of a branch node is computed so that more pure sub data bins can be obtained after separating the data using the rule. **C.** Using a one-hot encoder as shown allows the sklearn decision tree package to handle categorical data. **D.** Hold out test setup for checking the precision of rules.

**Origami Canopy (60% bending stiffness < 15000 N/m; 90% axial stiffness < 150000 N/m; 700 N/m < 90% bending stiffness < 1600 N/m;)**

Precision of kNN: 0.639; Precision of Boosting: 0.748

**ccp_alpha = 0.0001**

| tree number | 20 | 40 | 60 | 100 | 200 |
|---|---|---|---|---|---|
| depth=8 | 0.837 | 0.769 | 0.769 | 0.913 | 0.910 |
| depth=12 | 0.848 | 0.826 | 0.929 | 0.933 | 0.911 |
| depth=16 | 0.848 | 0.826 | 0.927 | 0.933 | 0.913 |
| depth=20 | 0.848 | 0.826 | 0.927 | 0.933 | 0.913 |
| depth=24 | 0.848 | 0.826 | 0.927 | 0.933 | 0.913 |
| depth=28 | 0.848 | 0.826 | 0.927 | 0.933 | 0.913 |
| depth=32 | 0.848 | 0.826 | 0.927 | 0.933 | 0.913 |

**ccp_alpha = 0.0005**

| tree number | 20 | 40 | 60 | 100 | 200 |
|---|---|---|---|---|---|
| depth=8 | 0.837 | 0.769 | 0.769 | 0.913 | 0.910 |
| depth=12 | 0.848 | 0.826 | 0.929 | 0.933 | 0.911 |
| depth=16 | 0.848 | 0.826 | 0.927 | 0.933 | 0.913 |
| depth=20 | 0.848 | 0.826 | 0.927 | 0.933 | 0.913 |
| depth=24 | 0.848 | 0.826 | 0.927 | 0.933 | 0.913 |
| depth=28 | 0.848 | 0.826 | 0.927 | 0.933 | 0.913 |
| depth=32 | 0.848 | 0.826 | 0.927 | 0.933 | 0.913 |

**ccp_alpha = 0.001**

| tree number | 20 | 40 | 60 | 100 | 200 |
|---|---|---|---|---|---|
| depth=8 | 0.837 | 0.769 | 0.769 | 0.913 | 0.910 |
| depth=12 | 0.848 | 0.826 | 0.929 | 0.933 | 0.911 |
| depth=16 | 0.848 | 0.826 | 0.927 | 0.933 | 0.913 |
| depth=20 | 0.848 | 0.826 | 0.927 | 0.933 | 0.913 |
| depth=24 | 0.848 | 0.826 | 0.927 | 0.933 | 0.913 |
| depth=28 | 0.848 | 0.826 | 0.927 | 0.933 | 0.913 |
| depth=32 | 0.848 | 0.826 | 0.927 | 0.933 | 0.913 |

**ccp_alpha = 0.002**

| tree number | 20 | 40 | 60 | 100 | 200 |
|---|---|---|---|---|---|
| depth=8 | 0.774 | 0.610 | 0.622 | 0.798 | 0.923 |
| depth=12 | 0.798 | 0.826 | 0.929 | 0.929 | 0.962 |
| depth=16 | 0.798 | 0.826 | 0.929 | 0.929 | 0.937 |
| depth=20 | 0.798 | 0.826 | 0.929 | 0.929 | 0.937 |
| depth=24 | 0.798 | 0.826 | 0.929 | 0.929 | 0.937 |
| depth=28 | 0.798 | 0.826 | 0.929 | 0.929 | 0.937 |
| depth=32 | 0.798 | 0.826 | 0.929 | 0.929 | 0.937 |

Precision value is average of 5 run with different random training set

**Fig. S6. Grid search results for the origami canopy dataset.** The table shows the mean precision of five separate runs for predicting the target class. The cost-complexity pruning alpha value does not have a significant influence on the precision.

**Origami Gripper (10Hz < frequency; heating power < 0.7W; maximum temperature < 500°C; 0.0015 N/m < stiffness)**
Precision of kNN: 0.413; Precision of Boosting: 0.753

**gini (minData=5)**

| tree number | 20 | 40 | 60 | 100 | 200 |
|---|---|---|---|---|---|
| depth=8 | 0.469 | 0.601 | 0.601 | 0.603 | 0.590 |
| depth=12 | 0.710 | 0.863 | 0.842 | 0.842 | 0.807 |
| depth=16 | 0.771 | 0.818 | 0.840 | 0.836 | 0.803 |
| depth=20 | 0.830 | 0.837 | 0.849 | 0.836 | 0.803 |
| depth=24 | 0.830 | 0.819 | 0.819 | 0.806 | 0.803 |
| depth=28 | 0.830 | 0.819 | 0.819 | 0.806 | 0.803 |
| depth=32 | 0.830 | 0.819 | 0.819 | 0.806 | 0.803 |

**entropy**

| tree number | 20 | 40 | 60 | 100 | 200 |
|---|---|---|---|---|---|
| depth=8 | 0.796 | 0.816 | 0.843 | 0.829 | 0.846 |
| depth=12 | 0.816 | 0.815 | 0.828 | 0.850 | 0.860 |
| depth=16 | 0.816 | 0.815 | 0.828 | 0.844 | 0.874 |
| depth=20 | 0.816 | 0.815 | 0.828 | 0.844 | 0.874 |
| depth=24 | 0.816 | 0.815 | 0.828 | 0.844 | 0.874 |
| depth=28 | 0.816 | 0.815 | 0.828 | 0.844 | 0.874 |
| depth=32 | 0.816 | 0.815 | 0.828 | 0.844 | 0.874 |

Precision value is average of 5 run with different random training set

**Fig. S7. Grid search results for the origami gripper dataset.** The table shows the mean precision of five separate runs for predicting the target class. The entropy criterion for selecting the splitting rule generally provides similar or better precision than using the Gini index.

**Origami Arch (Zstiff > 800 N/m; Xstiff > 600 N/m; error < 0.1;)**

Precision of kNN: 0.218; Precision of Boosting: 0.906

**subset ratio:** 50%

| tree number | 20 | 40 | 60 | 100 | 200 |
|---|---|---|---|---|---|
| depth=8 | 0.898 | 0.915 | 0.915 | 0.935 | 0.872 |
| depth=12 | 0.862 | 0.911 | 0.911 | 0.911 | 0.827 |
| depth=16 | 0.862 | 0.911 | 0.911 | 0.911 | 0.805 |
| depth=20 | 0.862 | 0.911 | 0.911 | 0.911 | 0.805 |
| depth=24 | 0.862 | 0.911 | 0.911 | 0.911 | 0.805 |
| depth=28 | 0.862 | 0.911 | 0.911 | 0.911 | 0.805 |
| depth=32 | 0.862 | 0.911 | 0.911 | 0.911 | 0.805 |

**subset ratio:** 60%

| tree number | 20 | 40 | 60 | 100 | 200 |
|---|---|---|---|---|---|
| depth=8 | 0.842 | 0.875 | 0.881 | 0.881 | 0.849 |
| depth=12 | 0.888 | 0.843 | 0.814 | 0.814 | 0.815 |
| depth=16 | 0.888 | 0.843 | 0.814 | 0.814 | 0.815 |
| depth=20 | 0.888 | 0.843 | 0.814 | 0.814 | 0.815 |
| depth=24 | 0.888 | 0.843 | 0.814 | 0.814 | 0.815 |
| depth=28 | 0.888 | 0.843 | 0.814 | 0.814 | 0.815 |
| depth=32 | 0.888 | 0.843 | 0.814 | 0.814 | 0.815 |

**subset ratio:** 70%

| tree number | 20 | 40 | 60 | 100 | 200 |
|---|---|---|---|---|---|
| depth=8 | 0.895 | 0.832 | 0.832 | 0.843 | 0.826 |
| depth=12 | 0.828 | 0.814 | 0.814 | 0.811 | 0.808 |
| depth=16 | 0.828 | 0.814 | 0.814 | 0.822 | 0.808 |
| depth=20 | 0.828 | 0.814 | 0.814 | 0.822 | 0.808 |
| depth=24 | 0.828 | 0.814 | 0.814 | 0.822 | 0.808 |
| depth=28 | 0.828 | 0.814 | 0.814 | 0.822 | 0.808 |
| depth=32 | 0.828 | 0.814 | 0.814 | 0.822 | 0.808 |

**subset ratio:** 80%

| tree number | 20 | 40 | 60 | 100 | 200 |
|---|---|---|---|---|---|
| depth=8 | 0.861 | 0.905 | 0.856 | 0.896 | 0.905 |
| depth=12 | 0.843 | 0.830 | 0.867 | 0.836 | 0.843 |
| depth=16 | 0.843 | 0.830 | 0.877 | 0.836 | 0.843 |
| depth=20 | 0.843 | 0.830 | 0.877 | 0.836 | 0.843 |
| depth=24 | 0.843 | 0.830 | 0.877 | 0.836 | 0.843 |
| depth=28 | 0.843 | 0.830 | 0.877 | 0.836 | 0.843 |
| depth=32 | 0.843 | 0.830 | 0.877 | 0.836 | 0.843 |

Precision value is average of 5 run with different random training set

**Fig. S8. Grid search results for the origami arch dataset.** The table shows the mean precision of five separate runs for predicting the target class. The subset ratio can affect the performance of the algorithm significantly, and the results show that a smaller number can give better results.

**Data S1 to S4. (Separate files)**

Three origami performance databases are published as supplementary material of this work. These three databases are:

- **Data S1:** Database on single Miura origami cell
- **Data S2:** Database on origami canopy
- **Data S3:** Database on active origami gripper
- **Data S4:** Database on origami arch

All three databases can be found in the electronic supplementary material associated with this manuscript or through the following GitHub website:

- https://github.com/zzhuyii/GenerateOrigamiDataSet

**Codes S1 to S2. (Separate files)**

Two sets of codes are published as supplementary material of this work. These two sets of codes are:

- **Codes S1** for simulating origami performances to populate the databases (Matlab)
- **Codes S2** for computing the decision rules using the decision tree algorithms (Python)

These two sets of codes can be found on GitHub using the following links:

- https://github.com/zzhuyii/GenerateOrigamiDataSet
- https://github.com/zzhuyii/TreeForOrigami